\documentstyle[11pt]{article}

\begin{document}

\title{ Observable and hidden singular features of
large fluctuations in nonequilibrium systems}
\author{ Mark I. Dykman$^{(1)}$\\
Department of Physics and Astronomy, Michigan State University,\\
East Lansing, MI 48824,\\
Mark M. Millonas\\
Theoretical Division, MS B258, Los Alamos National Laboratory,\\
Los Alamos, NM 87545,\\
and Vadim N. Smelyanskiy\\
Department of Physics, University of Michigan, Ann Arbor, MI 48104\\}

\maketitle

\begin{abstract}

We study local features, and provide a topological insight into the
global structure of the probability density distribution and of the
pattern of the optimal paths for large rare fluctuations away from a
stable state.  In contrast to extremal paths in quantum mechanics, the
optimal paths do {\it not} encounter caustics.  We show how this
occurs, and what, instead of caustics, are the experimentally
observable singularities of the pattern.  We reveal the possibility
for a caustic and a switching line to start at a saddle point, and
discuss the consequences.

\end{abstract}

The problem of large occasional fluctuations in nonequilibrium systems
is of substantial general interest and importance. These fluctuations
form the tails of statistical distribution, give rise to escape from a
stable state, and are responsible for the onset of many effects
investigated in various areas of physics --- some recent examples are
stochastic resonance\cite{SR} and transport in ratchets\cite{Man}. One
of the basic concepts in the analysis of large fluctuations is optimal
path --- the path along which the system moves, with overwhelming
probability, on its way to a given point remote from the stable state.
Optimal paths are experimentally observable, and have been an object
of active study for the last 20 years (see \cite{Wentzel,Lindenberg}
for a review). They play, in the context of fluctuations, the same
role as trajectories for dynamical systems, and therefore
understanding the pattern of the optimal paths is a key to
understanding large fluctuations.

 From the formal point of view, optimal paths are similar to the
extremal paths in quantum mechanics since both provide an extremum to
the integrands in the appropriate path integrals. A well-known feature
of the pattern of the extremal paths is the occurrence of caustics
\cite{Schulman}.  Caustics have also been revealed numerically in the
pattern of optimal paths for fluctuating systems of various types
\cite{Jauslin1} -- \cite{Ross}. For quantum mechanical
systems the physical meaning of caustics is well understood --- a
semiclassical wave function is oscillating on one side of a caustic
and exponentially decaying (or increasing) on the other side. In
contrast, the probability density distribution, which is determined by
the optimal paths, is nonnegative definite.  Therefore it cannot be
continued beyond a caustic, and it follows that caustics may {\it not}
be encountered by these paths.

In the present paper we address the problem of avoidance of caustics
by the physically meaningful optimal paths, and of the global
structure and the observable singularities of the pattern of these
paths. The physical and topological arguments we apply are quite
general, but as an illustration of how they work we consider the
simplest case, that of a two-variable system performing Brownian
motion described by the stochastic equation

\begin{equation}
\dot{q}_i = K_i({\bf q}) + \xi_i(t),\ \ \ {\bf q} \equiv
(q_1,\,q_2), \ \ \
\langle\xi_i(t)\rangle=0,
\end{equation}
\[ \langle\xi_i(t)\xi_j(t')\rangle=
D \delta_{ij}\delta(t-t'),\,\,\,\,i,j = 1,2. \]

\noindent Here, \mbox{\boldmath $\xi$}$(t)$ is Gaussian white
noise. The drift coefficients $K_{1,2}$ are assumed nonsingular
for finite ${\bf q}$.

We assume noise intensity $D$ to be small. In this case if the system
is prepared initially within the basin of attraction of an attractor
$a$, it will most likely approach the attractor in a characteristic
relaxation time $t_{\rm rel}$, as if there was no noise. Then it will
perform mostly small fluctuations about the position of the attractor
${\bf q}_a$, so that over $t_{\rm rel}$ a (quasi)stationary
probability density distribution $\rho_a({\bf q})$ will be formed.
Large fluctuations occasionally bring the system to points ${\bf q}$
remote from ${\bf q}_a$, and thus form the tails of $\rho_a({\bf q})$.
To logarithmic accuracy \cite{Wentzel}

\begin{equation}
\rho_a({\bf q}) = {\rm const} \times \exp(-S_a({\bf q})/D),
\end{equation}

\noindent
where $S_a({\bf q})$ is given by the solution of the variational
problem

\begin{equation}
S_a({\bf q}) = \min \int_{-\infty}^{0} {\cal L}(\dot{\bf
q}(t),{\bf q}(t))\ dt,
\end{equation}

\[{\cal L}(\dot{\bf q},{\bf q}) =
{1\over 2}(\dot {\bf q} - {\bf K})^2, \,\,\,
{\bf q}(-\infty) = {\bf q}_a,\,\,\, {\bf q}(0) =
{\bf q}.\]

\noindent Eq.(3) defines the optimal (most probable) path ${\bf
q}_{\rm opt}(t)$ to a point ${\bf q}$ from the stable position ${\bf
q}_a$ (${\bf K}({\bf q}_a) = 0$), in the small vicinity of which the
large fluctuation starts. The optimal path can be associated with the
trajectory of an auxiliary four-variable (two coordinates, $q_{1,2}$,
and two conjugate momenta, $p_{1,2}$) Hamiltonian system, with the
action $S_a({\bf q})$ and Lagrangian ${\cal L}$ (3), and with the
respective Hamiltonian ${\cal H} = {1\over 2} {\bf p}^2 + {\bf p\,K}$.
The Hamiltonian equations of motion for the trajectories are of the
form

\begin{equation}
\dot{\bf q} = {\bf K} + {\bf p}, \,\,\,\,
\dot{\bf p} = - ({\bf p\nabla}){\bf K} - {\bf p}\times({\bf
\nabla}\times{\bf K}),
\end{equation}

\noindent
where these trajectories lie on the energy surface  $E = 0$.

The approximation (2) is similar to the WKB approximation in quantum
mechanics, with the noise intensity $D$ corresponding to $i\hbar$. As
in quantum mechanics, the extremal paths ${\bf q}(t)$ (3),(4)
intersect each other, generically, and the set of these paths displays
caustics \cite{Jauslin1} -- \cite{Maier1}.

An interesting example of a system where caustics occur
\cite{Chinarov} is an underdamped nonlinear oscillator driven by a
nearly resonant force and by weak noise. Such an oscillator is a
reasonably good model of a few physical systems, including optically
bistable systems, and in particular a relativistic electron trapped in
a Penning trap and driven by cyclotron radiation \cite{Dehmelt}. We
emphasize that the onset of caustics in this system is not related to
bistability that emerges in a comparatively strong field and was
investigated in \cite{Chinarov}. This is seen from the pattern of
optimal paths shown in Fig.1. The variables $q_1, q_2$ are the
(dimensionless) coordinate and momentum of the oscillator in the
rotating frame. The equations of motion in this frame are of the form
(1) (cf. \cite{Chinarov}), with

\[K_1 =  q_2(q_1^2+q_2^2-1)-\eta\ q_1, \,\,
K_2 =  -q_1(q_1^2+q_2^2-1)-\eta\ q_2 +\sqrt{\beta}\]

\noindent
Here, $\eta$ is a dimensionless friction coefficient, and
$\beta^{1/2}$ is the dimensionless force amplitude (the bistability
arises for $\beta > \eta^2$).

It follows from the definition of the optimal path as the most
probable way to reach a given point that, if a point ${\bf q}$ can be
reached along two (or more) paths, only one of them is physically
meaningful: this is the path that provides an absolute minimum to the
action $S_a({\bf q})$. We show below that such a path has not ever
touched a caustic.

We first consider monostable systems, with the attractor $a$ being the
only steady state in the absence of noise. In this case the auxiliary
Hamiltonian system has only one fixed point (${\bf q} = {\bf q}_a,
{\bf p} = 0$). The trajectories $({\bf q}(t), {\bf p}(t))$ (4)
emanating from this point at $t \rightarrow - \infty$ form a smooth
flow (cf. \cite{Ludwig}) on a two-dimensional Lagrangian manifold (LM)
\cite{Arnold}.  Except for special cases (like detailed balance, ${\bf
\nabla}\times {\bf K} = 0$) the projections of LM onto the original
two-dimensional plane ($q_1, q_2$) will generally have singularities.

In two dimensions the only structurally stable types of singularities
\cite{Whitney} are folds and cusps, as illustrated in Fig. 2. The
projections of the folds of the LM are caustics. Each cusp gives rise
to a pair of folds, and in the case under consideration folds can only
begin or end at a cusp, or at infinity. This pattern is clearly seen
in the plot of the optimal paths of a periodically driven oscillator
in Fig.1,

It is a feature of the dynamics (3),(4) that the Lagrangian is
nonnegative definite. Thus, the action always increases along the
extremal paths. This corresponds, quite naturally, to a decrease in
probability density as the system moves along the path away from the
attractor.  An analysis which makes use of the normal form of the
action near a caustic \cite{Arnold} and of the explicit form of the
Hamiltonian ${\cal H}$ leads to an explicit local expression for the
action. From this it can be seen \cite{note} that the action to reach
a point along a path which {\it has not} touched a caustic {\it is
always less} than that along a path which {\it has passed} through a
caustic.

Near a cusp from which the caustics are going away
(a \lq\lq direction" of a caustic is  that of the paths for
which the caustic is an envelope) the probability distribution
can be obtained by modifying the appropriate results of the WKB
approximation in quantum mechanics\cite{Schulman}:

\[ \rho_a({\bf q}) \propto \exp\left(-{{\bf p}_c\cdot{\bf
q}\over D}\right) \int_{-\infty}^\infty  dP_1
\exp\left(\frac{\tilde
{S}(P_1,q_2)-P_1 q_1}{D}\right) ,\]
\begin{equation}
\tilde {S}(P_1,q_2)= -{1\over 4} a_{11} P_1^4 - {1\over
2} a_{12} P_1^2 q_2 - {1\over 2} a_{22} q_2^2.
\end{equation}

\noindent Here, $q_1,q_2$ are the coordinates measured from the
cusp point ${\bf q}_c = 0$ along the directions transverse to, and
parallel to the caustics at this point, i.e., the velocity of the path
in the cusp is pointing along $q_2$, ${\bf v}_c \equiv (\dot{\bf q})_c
= (0,v_c)$; ${\bf p}_c = - {\bf K}({\bf q}_c) + {\bf v}_c$, $a_{12} =
- v_c^{-1}$, $a_{22} =\partial \left(\left(2v_c\right)^{-1}{\bf K}^2 -
K_2\right)/ \partial q_2$ (the derivatives of ${\bf K}$ are evaluated
at the cusp point). The parameter $a_{11}$ depends on the global
features of the flow of the trajectories. It determines how sharply
the caustics $q_1 = \pm 2 a_{11}(q_2/3v_c)^{3/2}$ diverge with the
distance from the cusp $q_2$. The prefactor in the probability
distribution (2) blows up near the cusp point like $D^{-1/4}$.

For $|q_1|/D^{3/4}, |q_2|/D^{1/2}$ large and not close to the caustics
the integral (5) can be evaluated by the steepest descent method, and
the action $S_a({\bf q})$ in (2) can be expressed in terms of
$\tilde{S}(P_1, q_2)$ by implying $q_1 = \partial \tilde{S}/\partial
P_1$. For $q_2<0$ the action $S_a({\bf q})$ is single valued.  On the
other side of the cusp point, between the caustics, the action has
three values, that is, the surface $S_a({\bf q})$ has three sheets as
shown in Fig. 2(c).  The top sheet, the one with the largest $S_a$,
corresponds to the middle sheet of the LM in Fig. 2(a).  It is formed
by the paths which have been reflected from one of the caustics, and
contains the path which passes through the cusp point.  The two other
sheets of the surface $S_a$ and of the LM are formed by the paths
which have not touched a caustic.

Only the solution with the smallest $S_a({\bf q})$ should be kept in
(2) in the range of ${\bf q}$ where the distance between the sheets of
$S_a({\bf q})$ greatly exceeds $D$.  Therefore the top sheet of
$S_a({\bf q})$ is \lq\lq invisible" away from the cusp, and the
trajectories coming to the middle sheet of the LM in Fig.2 drop out of
the game. Two lower sheets of the action, $S_a^{(1)}$ and $S_a^{(2)}$,
which correspond to the lower and upper sheets of the LM, intersect
along a line with the projection $q_1^{(s)}(q_2)$ on the ${\bf
q}$-plane: $S_a^{(1)}(q_1^{(s)}, q_2)) = S_a^{(2)}(q_1^{(s)}, q_2)$.
This line starts at the cusp point and lies between the coalescing
caustics. There occurs switching at this line: the points a small
distance from each other, but lying on different sides of it are
reached along topologically different optimal paths ${\bf q}(t)$
(those providing $S_a^{(1)}$ or $S_a^{(2)}$). They approach the
switching line from the opposite sides.

The switching line can be immediately observed via experiments
\cite{Dykman} on the probability distribution of the paths ${\bf
q}(t)$ along which the system arrives to a given point. If this
distribution is measured for various positions of the final point, its
shape will change sharply once the final point crosses over the
switching line.  We notice that caustics may {\it not} be observed via
experiments of this sort, they are {\it hidden}: switching to another
path occurs {\it prior} a caustic being encountered.

The stationary distribution $\rho_a({\bf q})$ is regular in the
vicinity of a switching line: away from the cusp point it is
given by

\begin{equation}
\rho_a({\bf q}) = \sum_{i = 1,2} c^{(i)}({\bf q})\exp(-S_a^{(i)}({\bf
q})/D)
\end{equation}

\noindent
where the prefactors are evaluated for the paths lying on the
different sheets of the LM. However, the derivative of $D\ln \rho_a$
transverse to the switching line is discontinuous in the limit $D
\rightarrow 0$. This discontinuity was considered by Graham and
T\`{e}l \cite{Tel}, and by Jauslin \cite{Jauslin1} and Day
\cite{Day1}. The switching lines were found numerically in
\cite{Jauslin1}.

It is clear from the above picture that two switching lines emanating
from different cusp points can end in a point where they intersect
each other, and then another switching line starts at this point.
Therefore there arise physically observable trees of switching lines,
with the \lq\lq free" ends at cusp points.  Yet another conclusion
concerns the possibility, expected on physical grounds, to reach {\it
any point} $(q_1, q_2)$ along an optimal path which has never touched
a caustic. This possibility follows from the fact that caustics are
the only lines that limit the flow of the optimal paths ${\bf q}(t)$
in the considered case, and they emerge from the cusp points
simultaneously with the switching lines.  In particular the above
results provide an insight into the switching to a new escape path
observed in \cite{Maier1} when the old escape path is crossed by a
cusp point as a parameter is varied.

The structure of the singularities becomes more complicated if a
system has other steady states. A state of particular interest is an
unstable stationary state: a saddle point ${\bf q}_s$ (${\bf K(q}_s) =
0, {\rm det}[\partial K_i/\partial q_j] < 0$, and we assume ${\bf
\nabla\cdot K} < 0$). Saddle points occur on the basin boundaries
in multistable systems. In such systems, in addition to the
characteristic relaxation time $t_{\rm rel}$ of the deterministic
motion (that in the absence of noise), fluctuations about initially
occupied attractor $a$ are characterized by the reciprocal probability
$W_{a}^{-1}$ of the noise-induced escape from the basin of attraction.
In the time interval $t_{\rm rel} \ll t \ll W_{a}^{-1}$ the
probability distribution $\rho_a({\bf q})$ is quasistationary far from
the other attractors. We assume the basin boundary (the separatrix) to
extend to infinity and to contain only one unstable stationary point
${\bf q}_s$. It is the slowing down of the optimal path near ${\bf
q}_s$ that gives rise to the effects we discuss.

The point ${\bf q} = {\bf q}_s,\, {\bf p} = 0$ is a fixed point
of the Hamiltonian equations (4), and close to it they can be
linearized. We shall enumerate the eigenvectors \{${\bf
q}^{(n)},{\bf p}^{(n)}$\} ($n = 1,\ldots ,4$) so that the ones
with $n = 1,2$ are \lq\lq fluctuational", ${\bf p}^{(1,2)} \neq
0$, while the ones with $n = 3,4$ are \lq\lq deterministic",
${\bf p}^{(3,4)} \equiv 0$ (the solution or (4) with ${\bf p} =
0$ corresponds to the deterministic motion, $\dot{\bf q} = {\bf
K}$). In the vicinity of the fixed point

\begin{equation}
{\bf q}(t) = {\bf q}_s + \sum_{n = 1}^4 C^{(n)}
\exp(\lambda^{(n)}t){\bf q}^{(n)}
\end{equation}

\noindent
and similarly ${\bf p}(t) = \sum C^{(n)}
\exp(\lambda^{(n)}t){\bf p}^{(n)}$; $\lambda^{(1,2)}$ are the
eigenvalues of the matrix $-\partial K_i/\partial q_j$ evaluated in
the saddle point, and $\lambda^{(4,3)}$ are their negatives.  We
choose

\begin{equation}
\lambda^{(1)} > \lambda^{(3)} > 0, \,\,\lambda^{(2)} = -
\lambda^{(3)}, \, \lambda^{(4)} = - \lambda^{(1)}.
\end{equation}

The optimal path of particular importance is the one along which the
system escapes from the attractor. In a quite general case of a system
driven by Gaussian noise the most probable escape path
(MPEP) \cite{Maier2} ends up in the saddle point \cite{Lindenberg}.
Since MPEP approaches the saddle point as $t \rightarrow \infty$, for
this path $C^{1)} = C^{(3)} = 0$ in (7). The interrelation between the
coefficients $C^{(2)}$, $C^{(4)}$ is determined by the motion far away
from the saddle point (in special cases, like detailed balance,
$C^{(4)} = 0$). Because $|\lambda^{(4)}| > |\lambda^{(2)}|$, MPEP is
tangent to ${\bf q}^{(2)}$ in the saddle point (cf. \cite{Maier2}),
and for ${\bf q}$ lying on the MPEP

\begin{equation}
{\bf q}\times{\bf q}^{(2)} = M\,\left({\bf q}\cdot{\bf
q}^{(2)}\right)^{\lambda}\,{\bf q}^{(4)}\times{\bf
q}^{(2)},\ \ \lambda =
{\lambda^{(1)}\over\lambda^{(3)}}
\end{equation}

\noindent
(we chose the direction of ${\bf q}^{(2)}$ such that $C^{(2)} > 0$ for
the MPEP).

For the extremal paths other than MPEP $C^{(1,3)} \neq 0$. The
coefficients $C^{(1,3)}$ are interrelated via the expression
$C^{(1)}C^{(4)}/C^{(2)}C^{(3)} = r$ where the constant $r$ can be
found from the condition that the energy of the Hamiltonian motion $E
= 0$. The paths infinitesimally close to MPEP ($|C^{(1)}|
\rightarrow 0$) and lying on the opposite sides of it approach
asymptotically the eigenvectors $\pm {\bf q}^{(1)}$ as $t \rightarrow
\infty$ and then go away from the saddle point.  The corresponding
limiting paths form a \lq\lq cut" of the LM.  The singularities
related to the cut which are of central interest here have not been
considered in the analysis of the escape probability
\cite{Maier2,Schuss} where the absorbing boundary was placed along the
vector ${\bf q}^{(4)}$ (the basin boundary in the absence of
fluctuations).

If the cut was not crossed by other optimal paths emanating from a
given attractor $a$ it would determine the range that can be reached
from this attractor along an optimal path.  However, crossing of the
cut by the paths that have not encountered a caustic {\it may occur},
and in general, for $\lambda < 3/2$ a {\bf caustic emanates from the
saddle point} tangent to the cut. The equation for the caustic
$\partial (q_1, q_2)/\partial (t,\mu) = 0$ ($\mu$ is the parameter
that \lq\lq enumerates" the paths, the coordinate on a path ${\bf q}
\equiv {\bf q}(t,\mu)$) can be solved for small $C^{(1,3)}$
(but $|C^{(4)}| \exp\left(\lambda^{(2)}t\right) \ll
|C^{(1)}|\exp\left(\lambda^{(1)}t\right)$).
The resulting interrelation between the coordinates of
the caustic transverse and parallel to ${\bf q}^{(1)}$ is of the form:

\begin{equation}
{\bf q}\times{\bf q}^{(1)} = \frac{2 -
\lambda}{\lambda - 1}\left[\frac{\zeta}{
M}\,{\bf q}\cdot{\bf q}^{(1)}\right]^{\alpha}\,
\,{\bf q}^{(1)}\times{\bf
q}^{(2)},
\end{equation}

\noindent
where $ \alpha = 1/(2 - \lambda)$, ; $ 1 < \lambda < 3/2$, and

$$\zeta =
\lambda (1 - \lambda)\left({\bf p}^{(1)}\cdot{\bf
q}^{(4)}\right)\left({\bf q}^{(1)}\times{\bf
q}^{(3)}\right)/\left[\left({\bf p}^{(2)}\cdot{\bf
q}^{(3)}\right)\left({\bf q}^{(1)}\times{\bf
q}^{(2)}\right)\right]$$

Eq.(10) shows that the caustic is tangent to the cut in the saddle
point and is described by a simple power law with the {\it exponent}
$1 < \alpha < 2$ determined by {\it local} parameters of the motion
near the saddle point. The inequality $\alpha < 2$, or $\lambda <
{3\over 2}$ follows from the condition that the corrections to (10)
due to nonlinearity be small and gives the criterion for the onset of
the caustic \cite{note}. The prefactor in (10) depends on the nonlocal
characteristic $M$.

The general conclusion that optimal paths do not encounter caustics
applies to the caustic emanating from the saddle point as well. The
paths \lq\lq beating" the ones approaching the caustic emanate from
the saddle point themselves. When a system is approaching ${\bf q}_s$
its motion is slowed down, and it spends a time $\sim
\left(1/\lambda^{(1)}\right)|\ln D|$ performing small fluctuations
about ${\bf q}_s$. Over this time a large fluctuation can occur which
will drive the system away from ${\bf q}_s$. It is then necessary to
compare the probability to arrive to a given ${\bf q}$ directly from
the attractor or via ${\bf q}_s$, and it can be shown that the second
scenario wins on the caustic. The switching line emanates from ${\bf
q}_s$ and lies between the cut and the caustic. The system arrives on
opposite sides of it (as well as on the opposite sides of the cut on
the other side of the saddle point, cf. Fig.3) directly from the
attractor or having reached the saddle point first.

In conclusion, we have established the global structure of the pattern
of optimal paths for dissipative Markov systems and revealed singular
features related to the saddle points.  We have shown why and how
optimal paths avoid caustics. The singularities that can be
immediately observed experimentally by tracing optimal paths or by
measuring the probability distribution are switching lines. They start
at the cusp points from which caustics emanate or at the saddle, and
can form trees.

\noindent
$^{(1)}$ Present address: {\it Department of Physics,
Stanford University, Stanford, CA 94305}

\begin{figure}[t]
\vspace{6.5in}
\caption{\sf Pattern of optimal paths of a periodically
driven monostable nonlinear oscillator, $\eta = 0.1$ and $\beta =
0.0005$.}
\end{figure}

 \begin{figure}[t]
\vspace{6.5in}
\caption{\sf Generation of singularities.}
\end{figure}

\begin{figure}[t]
\vspace{6.5in}
\caption{\sf Saddle point}
\end{figure}

\end{document}